\documentstyle[graphics]{mn}

\newlength{\colwidth}
\newcommand{\ion}[2]{\hbox{#1\,{\sc #2}}}

\newcommand{\HI}{\ion{H}{i}}

\newcommand{\HeI}{\ion{He}{i}}
\newcommand{\HeII}{\ion{He}{ii}}
\newcommand{\HeIII}{\ion{He}{iii}}
\newcommand{\SiIV}{\ion{Si}{iv}}
\newcommand{\CIV}{\ion{C}{iv}}
\newcommand{\vpfit}{{\sc {VPFIT}}} 
\newcommand{\hydra}{{\sc {HYDRA}}}

\newcommand{\cm}{\rm cm}
\newcommand{\kms}{{\rm km}\,{\rm s}^{-1}}
\newcommand{\hmpc}{h^{-1}~{\rm Mpc}}

\newcommand{\lya}{Ly$\alpha$}
\newcommand{\lyb}{Ly$\beta$}

\title[The thermal history of the IGM]%
      {The thermal history of the intergalactic medium\thanks{ The
	data presented herein were obtained at the W.~M.~Keck
	Observatory, which is operated as a scientific partnership
	among the California Institute of Technology, the University
	of California and the National Aeronautics and Space
	Administration. The Observatory was made possible by the
	generous financial support of the W.~M.~Keck Foundation.}}
\author[J.~Schaye et al.] %
	{Joop~Schaye,$^1$ Tom~Theuns,$^{2}$ Michael~Rauch,$^{3}$
	George~Efstathiou$^1$ and \newauthor Wallace~L.W.~Sargent$^{4}$\\
	$^{1}$Institute of Astronomy, Madingley Road, Cambridge CB3
        0HA, UK\\
	$^{2}$Max-Planck-Institut f\"ur Astrophysik, Postfach 1523,
	85740 Garching, Germany\\
	$^{3}$European Southern Observatory,
        Karl-Schwarzschild-Str.\ 2, 85748 Garching, Germany\\
	$^{4}$Astronomy Department, California Institute of
        Technology, Pasadena, CA 91125, USA}

\begin{document}

\maketitle

\begin{abstract}
At redshifts $z \ga 2$, most of the baryons reside in the smooth
intergalactic medium which is responsible for the low column density
\lya\ forest. This photoheated gas follows a tight temperature-density
relation which introduces a cut-off in the distribution of widths of
the \lya\ absorption lines ($b$-parameters) as a function of column
density. We have measured this cut-off in a sample of nine high
resolution, high signal-to-noise quasar spectra, and determined the
thermal evolution of the intergalactic medium in the redshift range
2.0--4.5. At redshift $z \sim 3$, the temperature at the mean density
shows a peak and the gas becomes nearly isothermal. We interpret this
as evidence for the reionization of \HeII.
\end{abstract}


\section{INTRODUCTION}

According to the standard big bang model, the primordial hydrogen and
helium comprising the intergalactic medium (IGM) was hot and highly
ionized at early times. As the universe expanded, the hot plasma
cooled adiabatically, becoming almost completely neutral at a redshift
of $z \sim 10^3$. The IGM remained neutral until the first stars and
quasars began to produce ionizing photons. Eventually, the ionizing
radiation became intense enough to reionize hydrogen and later,
because of its higher ionization potential, to fully reionize
helium. Since the thermal evolution of the IGM depends strongly on its
reionization history, it can be used as a probe of the end of the
`dark ages' of cosmic history, when the first stars and quasars were
formed \cite{miralda-escude94:reionization,hui97:tempdens,%
haehnelt98:bparam}.

The absence of Gunn-Peterson absorption \cite{gunn65:gp_effect} in
quasar spectra, i.e.\ the complete absorption of quasar light blueward
of the \HI\ and \HeII\ \lya\ wavelengths requires that hydrogen must
have been highly ionized by $z\sim 5$
\cite{schneider91:gunn_peterson,songaila99:gunn_peterson} and helium
by $z\sim 2.5$ (Davidsen, Kriss \& Zheng 1996).  Measurements of the
\HeII\ \lya\ opacity suggest that helium may have reionized around $z \sim 3$
\cite{heap00:he,reimers97:he,jakobsen94:he,davidsen96:he,%
anderson99:he}. This would fit in with evidence for a hardening of the
UV background around this time, as derived from the ratio of
\SiIV/\CIV\ in high redshift quasar absorption lines
\cite{songaila96:metals,songaila98:ionizing_background}, although
both the observational result and its interpretation are still
controversial \cite{boksenberg98:ionizing_background,giroux97:siiv/civ}.

The resonant \lya\ absorption by residual low levels of neutral
hydrogen along the line of sight to a quasar produces a forest of
absorption lines.  Although many of the basic observational facts
about the \lya\ forest at high redshift ($z \sim 2$--5) had been
established before the 10\,m telescope era, the advent of the Keck
telescope has lead to much larger data samples at much higher
signal-to-noise ratio than hitherto available
\cite{hu95:lyaobs,lu96:lya_z=4,kirkman97:lya_obs}. The observational
progress has been matched on the theoretical side by semi-analytic
models \cite{bi97:lya_forest} and cosmological
hydro-simulations \cite{cen94:lya_forest,zhang95:lya_forest,%
petitjean95:lya_forest,hernquist96:lya_forest,miralda-escude96:lya_forest}
which together with the new data are now beginning to yield
significant quantitative cosmological constraints (see Rauch~1998 for
a recent review).

These simulations show that the low column density ($N \la
10^{14.5}\,\cm^{-2}$) absorption lines arise in a smoothly varying
IGM of low density contrast ($\delta \la 10$), which contains most of
the baryons in the universe. Since the overdensity is only mildly
non-linear, the physical 
processes governing this medium are well understood and relatively
easy to model. On large scales the dynamics are determined by gravity,
while on small scales gas pressure is important. Since shock heating
is unimportant for the low-density gas, the interplay between
photoionization heating and adiabatic cooling due to the expansion of
the universe results in a tight temperature-density relation, which is
well described by a power-law for densities around the cosmic mean,
$T=T_0(\rho/\bar{\rho})^{\gamma-1}$~\cite{hui97:tempdens}. This
relation is generally referred to as the `equation of state' (even
though the true equation of state is that of an ideal gas).

For models with abrupt reionization, the IGM becomes nearly isothermal
($\gamma \approx 1$) at the redshift of reionization. After
reionization, the temperature at the mean density ($T_0$) decreases
while the slope ($\gamma-1$) increases because higher density regions
undergo increased photoheating and expand less rapidly. Eventually, the
imprints of the reionization history are washed out and the equation
of state approaches an asymptotic state, $\gamma = 1.62$, $T_0 \propto
\left [ \Omega_bh^2 / \sqrt{\Omega_m h^2} \right ]^{1/1.7}$
\cite{miralda-escude94:reionization,hui97:tempdens,theuns98:apmsph}.
However, the timescale for recombination cooling in the low density
IGM is never small compared to the age of the universe for $z\la 20$
and inverse Compton cooling of free electrons off the cosmic microwave
background is only efficient for $z \ga 5$. Consequently, unless both
hydrogen and helium were fully reionized at redshifts considerably
higher than this, the gas will have retained some memory of when and
how it was reionized.

A standard way of analyzing \lya\ forest spectra is to decompose them
into a set of distinct absorption lines, assumed to have Voigt
profiles (e.g.\ Carswell et al.\ 1987). Various broadening mechanisms,
such as Hubble broadening (the differential Hubble flow across the
absorber), peculiar and thermal velocities contribute to the line
widths (Meiksin 1994; Hui \& Rutledge 1999; Theuns, Schaye \& Haehnelt
2000). However, there exists a lower limit to the line-width, set by
the temperature of the gas. Because the physical density of the IGM
correlates strongly with the column density of the absorption lines,
this results in a cut-off in the distribution of line widths
($b$-parameters) as a function of column density, which traces the
equation of state of the gas (Schaye et al.\ 1999, hereafter STLE;
Ricotti, Gnedin \& Shull 2000; Bryan \& Machacek 2000). Hence we can
infer the equation of state of the IGM by measuring the minimum \lya\
line width as a function of column density.

Here, we measure the $b(N)$ cut-off in nine high resolution, high S/N
quasar spectra, spanning the redshift range 2.0--4.5. We use
hydrodynamic simulations to calibrate the relations between the
parameters of the $b(N)$ cut-off and the equation of state. By
applying these relations to the observations, we are able to measure
the evolution of the equation of state over the observed redshift
range. We find that the thermal evolution of the IGM is drastically
different from that predicted by current models. The temperature peaks
at $z\sim 3$, which, together with supporting evidence from
measurements of the \HeII\ opacity and the \SiIV/\CIV\ ratios, we
interpret as evidence for the second reionization of helium (\HeII\
$\rightarrow$ \HeIII). Ricotti et al.~\shortcite{ricotti00:eos}
recently applied a similar technique to published lists of Voigt
profile fits. A comparison with the method and results of Ricotti et
al.\ is given in section~\ref{sec:discussion}.

This paper is organized as follows. In sections \ref{sec:observations}
and \ref{sec:simulations} we describe the observations and the
simulations respectively. We discuss the difference between evolution
of the $b$-distribution and evolution of the temperature in
section~\ref{sec:b-evolution}. In section~\ref{sec:method} we briefly
describe our method for measuring the equation of state, before we
present our results in 
section~\ref{sec:results}. Systematic errors are discussed in
section~\ref{sec:systematics}. Finally, we discuss and summarize the main
results in section \ref{sec:discussion}.

\section{OBSERVATIONS}
\label{sec:observations}

We analyzed a sample of nine quasar spectra, spanning the redshift
range $z_{\rm em}=2.14$--4.55 (Table~\ref{tbl:qsos}). The spectra of
Q1100$-$264 and APM~08279+5255 were kindly provided by R.~Carswell and
S.~Ellison respectively. All spectra were taken with the
high-resolution spectrograph (HIRES, Vogt et al.\ 1994) on the Keck
telescope, except the spectrum of Q1100$-$264, which was taken with
the UCL echelle spectrograph of the Anglo Australian
Telescope. Details on the data and reduction procedures, as well as
the continuum fitting, can be found in Carswell et
al.~\shortcite{carswell91:1100} for Q1100$-$264, Ellison et
al.~\shortcite{ellison99:0827} for APM~08279+5255 and Barlow \&
Sargent~\shortcite{barlow97:reduction} and Rauch et
al.~\shortcite{rauch97:lya_opacity} for the others.  The nominal
velocity resolution (FWHM) was $8\,\kms$ for Q1100$-$264 and
$6.6\,\kms$ for the others and the data were rebinned onto $0.04\,{\rm
\AA}$ pixels on a linear wavelength scale. The signal to noise ratio
per pixel is typically about 50, except for Q1100$-$264 for which it
is about 20.
\begin{center}
\begin{table}
\caption{Quasar spectra used}
\begin{tabular}{ll}
QSO & $z_{\rm em}$\\
\hline
Q1100$-$264 & 2.14\\
Q2343+123 & 2.52\\
Q1442+293 & 2.67\\
Q1107+485 & 3.00\\
Q1425+604 & 3.20\\
Q1422+231 & 3.62\\
APM~08279+5255 & 3.91\\
Q0000$-$262 & 4.11\\
Q2237$-$061 & 4.55\\
\hline
\end{tabular}
\label{tbl:qsos}
\end{table}
\end{center}


In order to avoid confusion with the \lyb\ forest, only the region of
a spectrum between the quasar's \lyb\ and \lya\ emission lines was
considered. In addition, spectral regions close to the quasar
(typically 8--10$\,\hmpc$, but $21\,\hmpc$ for APM~08279+5255 and
$32\,\hmpc$ for Q1100$-$264) were omitted to avoid proximity
effects. Regions thought to be contaminated by metals and damped \lya\
lines were removed (metal line regions were identified by correlating
with metal lines redwards of the quasar's Ly-alpha emission line and
with strong \HI\ lines). The absorption features in the remaining
spectral regions were fitted with Voigt profiles using the same
automated version of \vpfit~\cite{webb87:thesis,carswell87:vpfit} as
was used for the simulated spectra. Using a fully automatic fitting
program invariably results in a few `bad fits'. However, given that
there is no unique way of decomposing intrinsically non-Voigt
absorption lines into a set of discrete Voigt profiles, it is
essential to apply the same algorithm to simulated and observed
spectra. Since `bad fits' will also occur in the synthetic spectra, we
have made no attempt to correct them.

The \lya\ forest of a single quasar spans a considerable redshift
range ($\Delta z \sim 0.5$). In order to minimize the effects of
redshift evolution and S/N variation across a single spectrum, we
divided each \lya\ forest spectrum into two parts of equal length. STLE
showed that their algorithm for measuring the cut-off of the $b(N)$
distribution is relatively insensitive to the number of absorption
lines, the statistical variance is almost the same for e.g.\ 150 and
300 lines. Hence little information is lost by analyzing narrow
redshift bins if the absorption line density is high. The two halves
of the spectra were analyzed separately and each was compared with its
own set of simulated spectra (see section~\ref{sec:simulations}). For
the two lowest redshift quasars the 
number of absorption lines is too small to split the data in
half. Hence each quasar, except for Q1100$-$264 and Q2343+123, provides
two nearly independent data sets.

The complete set of absorption line samples is listed in
Table~\ref{tbl:samples}. The median, minimum and maximum redshifts of
the absorption lines used to determine the $b(N)$ cut-off are listed in
columns 2--4. The minimum column density considered was set to
$10^{12.5}\,\cm^{-2}$ for all samples, since blends dominate at lower
column densities. The maximum column densities are listed in column 5,
they were determined by the following considerations: (a) the cut-off 
can be measured more accurately if the column density interval is larger;
(b) we only want to measure the cut-off for column densities that
correspond to the density range for which the gas follows a power-law
temperature-density relation. The total number of absorption lines in
each sample is listed in column 6 (only lines for which \vpfit\ gives
relative errors in both the $b$-parameter and column density less than
0.25 are considered). Finally, column 6 lists the pivot column density
of the power-law cut-off, $b=b_{N_0}(N/N_0)^{\Gamma-1}$, that was fit
to the data (c.f.~section~\ref{sec:method}).
\begin{center}
\begin{table}
\caption{Observed absorption line samples. Columns 2--4 correspond to
the median, minimum and maximum redshifts of the absorption lines in
the sample. Column~5 is the maximum column density considered (the
minimum column density is always $10^{12.5}\,\cm^{-2}$) and column~6
contains the total number of absorption lines used to determine the
$b(N)$ cut-off. The last column is the pivot column density of the
power-law cut-off, $b=b_{N_0}(N/N_0)^{\Gamma-1}$.} 
\begin{tabular}{lccccrcc}
Sample & $z$ & $z_{\rm min}$ & $z_{\rm max}$ & $\log N_{\rm max}$ & \# lines &
$\log N_0$ \\
\hline
1100  & 1.96 & 1.85 & 2.09 & 14.0 & 44  & 13.0 \\
2343  & 2.29 & 2.23 & 2.38 & 14.2 & 48  & 13.0 \\
1442a & 2.21 & 2.10 & 2.33 & 14.5 & 87  & 13.0 \\
1442b & 2.50 & 2.33 & 2.63 & 14.5 & 90  & 13.0 \\
1107a & 2.59 & 2.48 & 2.71 & 14.5 & 76  & 13.0 \\
1107b & 2.84 & 2.71 & 2.95 & 14.5 & 91  & 13.0 \\
1425a & 2.66 & 2.55 & 2.88 & 14.5 & 94  & 13.0 \\
1425b & 3.00 & 2.89 & 3.14 & 14.5 & 118 & 13.0 \\
1422a & 3.08 & 2.91 & 3.22 & 14.5 & 145 & 13.0 \\
1422b & 3.37 & 3.22 & 3.53 & 14.5 & 152 & 13.0 \\
0827a & 3.23 & 3.15 & 3.42 & 14.5 & 105 & 13.4 \\
0827b & 3.55 & 3.43 & 3.70 & 14.5 & 106 & 13.4 \\
0000a & 3.72 & 3.61 & 3.81 & 14.8 & 77  & 13.5 \\
0000b & 3.91 & 3.81 & 4.01 & 14.8 & 97  & 13.5 \\
2237a & 3.84 & 3.69 & 4.02 & 14.8 & 140 & 13.5 \\
2237b & 4.31 & 4.15 & 4.43 & 14.8 & 127 & 13.5 \\
\hline
\end{tabular}
\label{tbl:samples}
\end{table}
\end{center}

\begin{figure*}
\resizebox{\textwidth}{!}{\includegraphics{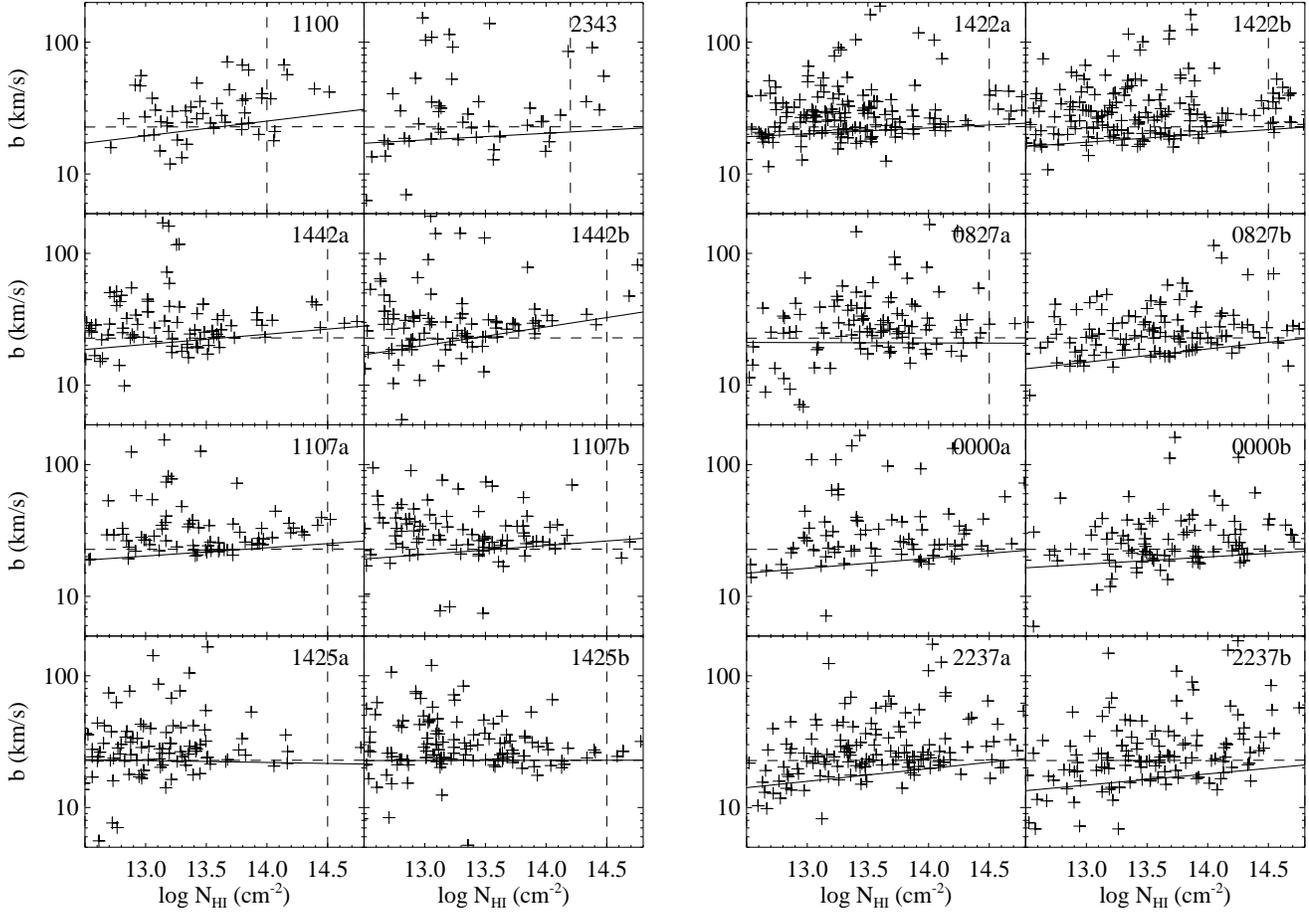}}
\caption{$b(N)$-Distributions for the observed samples listed in
Table~\protect\ref{tbl:samples}. Crosses indicate positions of
absorption lines, errors are not displayed. Only lines that are used
for the determination of the cut-off are shown, i.e.\ lines for which
\vpfit\ gives relative errors in both $b$ and $N$ smaller than 25 per
cent. Solid lines are the measured cut-offs. Vertical dashed lines
indicate the maximum column density used when fitting the
cut-off. Horizontal dashed lines are identical and correspond to the
cut-off of sample 1425b. Note that evolution of the $b(N)$ cut-off
cannot be interpreted as evolution of the equation of state in any
straightforward way, because the density-column density relation
changes with redshift (see
section~\ref{sec:b-evolution}). Furthermore, the cut-offs in different 
panels can only be compared directly if they contain a similar
number of absorption lines.} 
\label{fig:mosaic}
\end{figure*}

Scatter plots of the $b(N)$-distribution for all observed samples
(Table~\ref{tbl:samples}) are shown in Fig.~\ref{fig:mosaic}. The
solid lines are the measured cut-offs, the vertical dashed lines
indicate the maximum column density used for fitting the
cut-off. There are clear differences between the samples. We will show
in section~\ref{sec:b-evolution} that even a non-evolving
$b$-distribution would imply a strong thermal evolution. Several
samples contain a few lines that fall far below the 
cut-off. These lines, which have no significant effect on the measured
cut-off, are most likely blends or unidentified metal lines.

\begin{figure}
\resizebox{\colwidth}{!}{\includegraphics{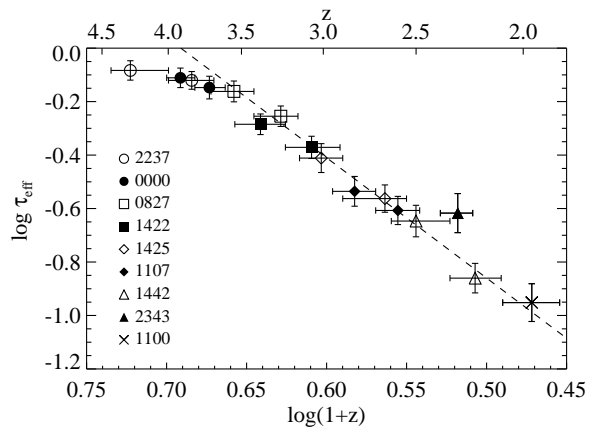}}
\caption{The observed effective optical depth as a function of
redshift. The dashed line, which has a slope of 4.5, is drawn to guide
the eye. Horizontal error bars indicate redshift intervals used,
vertical error bars are 1$\,\sigma$ errors as determined from a
bootstrap analysis using chunks of $4\,$\AA.}
\label{fig:tau_eff}
\end{figure}

Fig.~\ref{fig:tau_eff} shows the effective optical depth, $\tau_{\rm
eff} \equiv -\ln(\left < F \right >)$, of the observed samples as a
function of (decreasing) redshift. The ionizing background in the
simulations was rescaled to match these effective optical depths. The
scatter is small, considering that most data points represent just
half of a \lya\ forest spectrum. Note that Rauch et
al.~\shortcite{rauch97:lya_opacity} studied the opacity of the forest
using seven out of nine of the quasars from this sample. They found a
slightly less rapid increase with redshift, because they rebinned the
data into three redshift bins, centered on $z=2$, 3 and 4.

\section{SIMULATIONS}
\label{sec:simulations}

In order to calibrate the relation between the parameters of the
$b(N)$ cut-off and the effective equation of state, we have simulated
eight variants of the 
currently favoured flat, scale invariant, cosmological constant
dominated cold dark matter model, which vary only in their heating
rates (Table~\ref{tbl:models}). The calibration was repeated for each of the
observed samples of absorption lines listed in
Table~\ref{tbl:samples}. Synthetic spectra were computed along 
1200 random lines of sight through the simulation box at the nearest
redshift output ($\Delta z=0.25$). The background flux was rescaled
such that the mean effective optical depth in the simulated spectra
matches that of the observed sample. Each spectrum was convolved with
a Gaussian with full width at half maximum (FWHM) identical to that of
the observations and resampled onto pixels of the same size. The noise
properties of the observed spectrum were computed as a function of
flux and imposed on the simulated spectra. The resulting spectra were
continuum fitted as described in Theuns et
al.~\shortcite{theuns98:apmsph}. Finally, Voigt profiles were fitted
using the same automated version of \vpfit\ as was used for the
observed spectra. We will refer to the sample of lines drawn from the
synthetic spectra of simulation X, designed to mimic the observed
spectrum Y as as model X-Y, e.g.\ model L1-1442a.

All models have a total matter density
$\Omega_m=0.3$, vacuum energy density $\Omega_\Lambda=0.7$, baryon
density $\Omega_bh^2=0.019$, present day Hubble constant
$H_0=65\,{\kms}\,{\rm Mpc}^{-1}$ and the amplitude of the initial
power spectrum is normalized to $\sigma_8=0.9$. The IGM is assumed to be of
primordial composition with a helium abundance of 0.24 by mass and is
photoionized and photoheated by the UV-background from quasars.
The exact cosmology and UV-background is unimportant for this
analysis, as is the normalization (see
section~\ref{sec:systematics}). Although these parameters may affect
the equation of state of the IGM in the simulations, they do not
change its relation to the $b(N)$ cut-off.
\begin{center} 
\begin{table}
\caption{Simulations used for calibrating the relation between
the $b(N)$ cut-off and the effective equation of state.}
\begin{tabular}{lccl}
Model & $\epsilon_{\rm He}$ & $\mathcal{H}_{\rm X}$ & Comment\\ 
\hline
L0.3  & 1/3 & 0 & low $T_0$ \\
L1    & 1   & 0 & reference model \\
L2    & 2   & 0 & high $T_0$ \\
L3    & 3   & 0 & very high $T_0$ \\
Lx    & 1   & 1 & low $\gamma$, high $T_0$ \\
Lx2.5 & 1   & 5/2 & very low $\gamma$, high $T_0$\\
Lx5   & 1   & 5 & very low $\gamma$, very high $T_0$\\
L1e   & 1   & 0 & high $\gamma$ \\
\hline
\end{tabular}
\label{tbl:models}
\end{table}
\end{center}

The numerical simulations used in this paper follow the evolution of a
periodic, cubic region of the universe and are performed with a
modified version of \hydra~\cite{couchman95:hydra}, which uses smooth
particle hydrodynamics~\cite{lucy77:sph,gingold77:sph}. The
simulations employ $64^3$ gas particles and $64^3$ cold dark matter
particles in a box of of comoving size 3.85~Mpc, so the particle mass
is $1.14\times 10^6\,{\rm M}_\odot$ for the gas and $6.51\times
10^6\,{\rm M}_\odot$ for the dark matter.

Our reference model, L1, is photoionized and photoheated by the
UV-background from quasars as computed by Haardt \& Madau~(1996,
hereafter HM), using the optically thin
limit. Models L0.3, L2 and L3 are identical, except that we have
multiplied the helium photoheating rates (column $\epsilon_{\rm He}$
in Table~\ref{tbl:models}) by factors of 1/3, 2 and 3
respectively (keeping the ionization rates constant). The effective
helium photoheating rate may be higher 
than computed in the optically thin limit because of radiative
transfer effects (e.g.\ Abel \& Haehnelt 1999). Model Lx is identical
to model L1, except that we have included Compton 
heating by the hard X-ray background as computed by Madau \&
Efstathiou~\shortcite{madau99:xray} (column $\mathcal{H}_{\rm X}$ in
Table~\ref{tbl:models}). For a highly ionized plasma, the
energy input per particle from Compton scattering of free electrons is
independent of the density. Hence, Compton heating tends to flatten
the effective equation of state. We have used this fact to
artificially construct models with low values of $\gamma$, by
multiplying the X-ray heating rates by (unrealistic) factors of 2.5
and 5 for models Lx2.5 and Lx5 respectively. Finally, model L1e is
identical to model L1, except that we have set the ionization and
heating rates for \HI\ and \HeI\ for redshifts between 6 and 10 equal
to those at $z=6$. In this model \HI\ and \HeI\ ionize early (at
$z=10$), which drives $\gamma$ to larger values.

In addition to the models listed in Table~\ref{tbl:models}, we have
performed some simulations to investigate possible systematic
effects. We simulated model L1 twice with lower resolutions ($2\times
54^3$ and $2\times 44^3$ particles) and model L3 in a larger box, but
with the same resolution ($5\,\hmpc$ and $2\times 128^3$ particles
instead of $2.5\,\hmpc$ and $2\times 64^3$ particles). Model L1 was
simulated twice more with lower normalizations of the initial power
spectrum ($\sigma_8=0.65$ and 0.4 instead of 0.9).

\section{Evolution of the \lowercase{$b$}-distribution vs thermal evolution}
\label{sec:b-evolution}

Williger et al.~\shortcite{williger94:lya} and Lu et
al.~\shortcite{lu96:lya_z=4} found that the $b$-parameters at $z\sim
4$ are smaller than at $z=2$--3. Kim et
al.~\shortcite{kim97:lya_evolution} showed that the increase in the
line widths with decreasing redshift continues over the range $z=3.5$
to 2.1. It is tempting to interpret these results as evidence for
an increase in the temperature $T_0$ with decreasing
redshift. However, we will show in this section that the $b$-values
are smaller at higher redshift even for models in which $T_0$ is
higher, as is the case for models in which the universe is fully
reionized by $z=4$. 

\begin{figure}
\resizebox{\colwidth}{!}{\includegraphics{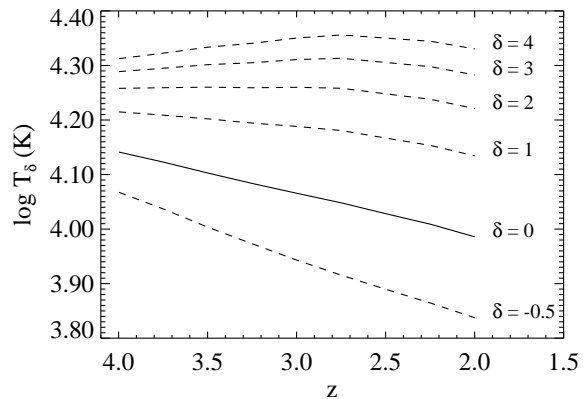}}
\caption{The temperature at a given density contrast as a function of
redshift for model~L1, which uses the HM ionizing background. Although
the temperature at the mean density, $T_0$, increases with redshift (solid
line), the temperature at slight overdensities is almost constant, or
even decreasing with redshift.}
\label{fig:L1-eos}
\end{figure}

As pointed out by STLE, any statistic that is sensitive to the
temperature of the absorbing gas, will in general depend on both the
amplitude, $T_0$, and the slope, $\gamma$, of the equation of
state. This is because temperature is a function of density and the
absorbing gas is, in general, not all at the mean density of the
universe.  After reionization, $T_0$ will decrease and $\gamma$ will
increase with time. Consequently, the evolution of the temperature at
a given overdensity can be very different from the evolution of
$T_0$. This is illustrated in Fig.~\ref{fig:L1-eos}, where the
temperature $T_\delta$ at a density contrast $\delta \equiv
\rho/\bar{\rho}-1$ is plotted as a function of redshift for
model~L1. Even though the temperature at the mean density decreases with
time (solid line), the temperature at a density contrast as little as
2 remains almost constant.

\begin{figure}
\resizebox{\colwidth}{!}{\includegraphics{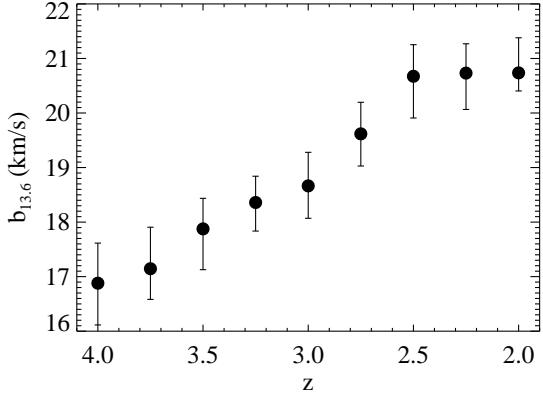}}
\caption{The $b(N)$ cut-off at a fixed column density
($10^{13.6}\,\cm^{-2}$) as a function of redshift for model~L1, which
uses the HM ionizing background. The error bars enclose 68 per cent
confidence intervals around the medians, as determined from 1000 sets
of 300 absorption lines. The cut-off shifts to larger $b$-values with
decreasing redshift, even though the temperature at the mean density,
$T_0$, decreases (solid line in Fig.~\ref{fig:L1-eos}).}
\label{fig:L1-b0}
\end{figure}

The general expansion of the universe ensures that the column density
corresponding to a fixed overdensity is a strongly increasing function
of redshift. In fact, most of the evolution of the \lya\ forest can be
understood in terms of the resulting scaling of the optical depth
(e.g.\ Hernquist et al.~1996; Dav\'e et al.~1999; Machacek et
al.~1999).  When interpreting the evolution of the $b$-distribution,
one therefore has to keep in mind that: (a) at fixed column density,
absorption lines at higher redshift will correspond to absorbers of
smaller overdensities; (b) the evolution of the temperature at a fixed
overdensity depends on the evolution of both $T_0$ and
$\gamma$. Together these effects can conspire to make the
$b$-parameters smaller at higher redshift, even when the temperature
$T_0$ is higher. Fig.~\ref{fig:L1-b0} shows that this will happen for
models in which the IGM is fully reionized at the observed redshifts
($z\la 4$).

The discussion in this section shows that to derive the evolution of
$T_0$ using a statistic that is sensitive to the temperature of the
absorbing gas, one needs to determine the evolution of: (1) the
temperature of the gas; (2) the overdensity of the gas and (3) the
slope of the equation of state. We will see that the uncertainty in
$\gamma$ is the limiting factor.

\section{Method}
\label{sec:method}

STLE demonstrated that the observed cut-off in the distribution of
$b$-parameters as a function of column density can be used to measure
the equation of state of the IGM. In particular, they showed that the
$b(N)$ cut-off can be fitted by a power-law,
$b=b_{N_0}(N/N_0)^{\Gamma-1}$, whose parameters $\log b_{N_0}$ and
$\Gamma-1$ are proportional to $\log T_{\delta (N_0)}$ and $\gamma-1$
respectively. For each observed sample, we first calibrate these
relations using the simulations and then use them to convert the
observed cut-offs into measurements of the equation of state.

For each observed sample of absorption lines (Table~\ref{tbl:samples})
we go through the following procedure. First mock spectra are
generated from the 8 simulations listed in
Table~\ref{tbl:models}. The synthetic spectra are processed to give
them the same characteristics (mean absorption, resolution, pixel size
and noise properties) as the corresponding observed spectra. These are
then fitted with Voigt profiles using the same automated fitting
package that was used for the observations. For each of the eight
simulated sets of absorption lines, the $b(N)$ cut-off is
fitted using the iterative procedure developed by STLE.
We then use these 8 simulations to calibrate the relations between the
parameters of the $b(N)$ cut-off, $(b_{N_0},\Gamma)$ and the
parameters of the equation of state, $(T_{\delta(N_0)},\gamma)$. 
We measure the density contrast corresponding to the pivot column density,
$N_0$, by using the fact $\log T_{\delta(N_0)} \propto \log b_{N_0}$
(STLE), essentially because both thermal broadening and Jeans
smoothing scale as the square root of the temperature. 

\begin{figure*}
\resizebox{0.9\textwidth}{!}{\includegraphics{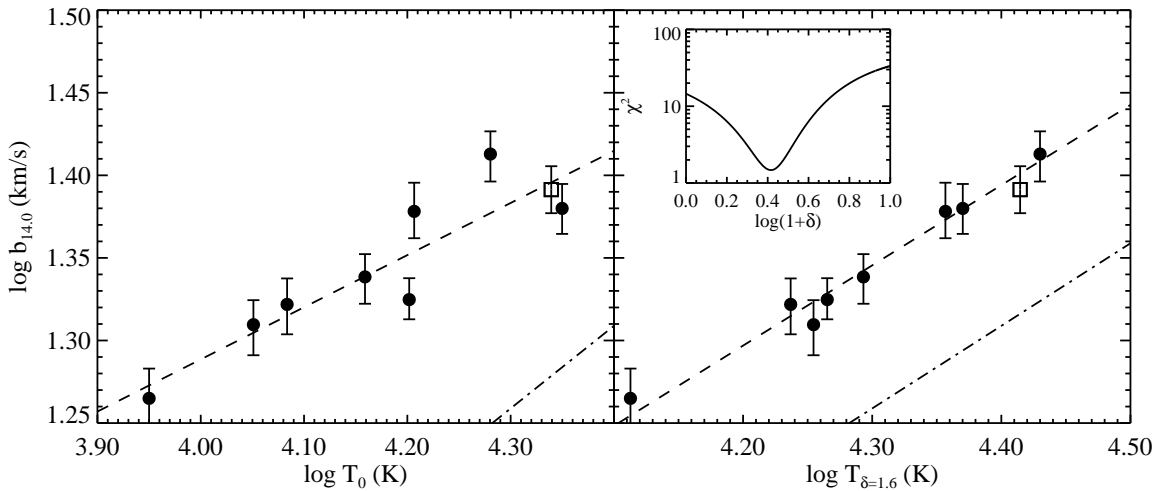}}
\caption{The intercept of the $b(N)$ cut-off as a function of the
temperature at the mean density (left panel) and at a density contrast
of $\delta=1.6$ (right panel). Filled circles are the simulations
corresponding to the observed sample 1422a. The open square is for a
simulation whose thermal evolution matches the observations (dashed
line in Fig.~\ref{fig:results}). The dashed line is the least-squares
fit for the filled circles. The dot-dashed line indicates the
$b$-value corresponding to pure thermal broadening:
$b=(2k_BT/m_p)^{1/2}$. The inset in the right panel shows how the
(total) $\chi^2$ of the fit varies as a function of $\delta$; $\chi^2$
is minimum for $\delta=1.6$, which is the density contrast
corresponding to the column density of the intercept of the $b(N)$
cut-off. ($10^{14.0}\,\cm^{-2}$ in this example). The difference
between the dashed and the dot-dashed lines in the right panel is due
to the contribution of baryon smoothing to the widths of the lines
around the $b(N)$ cut-off.}
\label{fig:intercept-logt}
\end{figure*}

Fig.~\ref{fig:intercept-logt} illustrates our method for measuring
$T_\delta$ and $\delta(N_0)$. In the
left panel the intercept of the cut-off, measured at
$N_0=10^{14.0}\,\cm^{-2}$ in this example, is plotted as a function of
$\log T_0$ for the simulations of sample 1422a (filled circles). As
expected, the relation between the intercept, $\log b_{14.0}$, and $\log T_0$
is not very tight. The large scatter arises because $\log b_{14.0}$ is
not proportional to $\log T_0$, but to $\log T_{\delta(N_0)}$,
where $\delta(N_0)$ is the density contrast corresponding to the pivot column
density $N_0=10^{14.0}\,\cm^{-2}$. Indeed, if we plot $\log b_{14.0}$ as
a function of $\log T_{\delta=1.6}$ (right panel), the scatter becomes
very small, implying that $\delta(N_0)\approx 1.6$.

Changing the value of $\delta$ shifts the data points in the plot
horizontally, the amount depending on the value of $\gamma$ in the
simulation corresponding to the data point. Because we do not know a
priori what the density contrast corresponding to $N_0$ is, we vary
$\delta$ and see for which value the scatter is minimal. The inset in
the right panel shows the total $\chi^2$ of the linear least squares
fit to the data points as a function of the density contrast. For this
example the scatter is minimal for $\delta=1.6$, which is the density
contrast used in the right panel. Using
the optically depth weighted density - column density relation,
introduced by STLE, we
find that the column density $N=10^{14.0}\,\cm^2$ does indeed
correspond to a density contrast of about 1.6.

The dot-dashed lines in Fig.~\ref{fig:intercept-logt} show the
$b$-value expected for pure thermal broadening, $b=\sqrt{2 k_B
T/m_p}$, where $k_B$ is the Boltzmann constant and $m_p$ is the proton
mass. The relation between $\log b_{14.0}$ and $\log T_{\delta=1.6}$
lies slightly above the pure thermal broadening line, but has the same
slope. This implies that the widths at the cut-off are dominated by
thermal broadening, with an additional component that also scales as
$T^{1/2}$. We identify this last component as the differential Hubble
flow across the absorber, whose size is set by the Jeans smoothing
scale and does indeed depend on the square root of the temperature. We
consistently found that minimizing the scatter in the $\log
b_{N_0}$-$\log T_\delta$ relation by varying $\delta$ results in a
relation that has a slope of about 0.5 and that the value of $\delta$
found agrees well with direct measurements of the density contrast
corresponding to the column density $N_0$. We therefore use this
procedure to estimate $\delta$ and $\log T_\delta$ and conservatively
estimate the error in the density contrast to be
$\sigma(\log(1+\delta))=0.15$.  Although for this work the error from
the determination of the density - column density relation does not
contribute significantly to the total error in $T_0$, it may well
become the limiting factor when a larger sample of quasars is used.

Having measured $T_\delta$, $\delta$ and $\gamma$ we can compute
$T_0$,
\begin{equation}
\log T_0 = \log T_\delta - (\gamma-1)\log(1+\delta).
\label{eq:logt0}
\end{equation}
Each measurement of $\log b_{N_0}$ and $\Gamma$ comes with associated
errors, which are determined from the bootstrap distribution (c.f.\
STLE). These errors can be converted directly into errors in $\log
T_\delta$ and $\gamma$ respectively, using the linear relations
between the cut-off and the equation of state, determined from the
simulations. To these errors we add in quadrature the residual scatter
of the data points around the linear fit. The error in $\log T_0$ is
then given by,
\begin{eqnarray}
\Delta^2 (\log T_0) & = & \Delta^2 (\log T_\delta) + \left
	[\Delta(\gamma)\log(1+\delta) \right ]^2 + \nonumber \\ & &
	\left [ (\gamma-1)\Delta(\log(1+\delta)) \right ]^2.
\end{eqnarray}

After having measured the thermal evolution of the IGM, we ran a
simulation designed to match the observations (dashed lines in
Fig.~\ref{fig:results}). The open square in
Fig.~\ref{fig:intercept-logt} corresponds to this simulation. The
difference between the evolution in the calibrating simulations and
the true evolution is particularly large at $z\sim 3$ (c.f.\
Fig.~\ref{fig:results}a), the redshift of model 1422a. The fact that
the open square follows the same relation as the other data points,
confirms that a model whose equation of state matches the one
determined from the observations using the methods described in this
section, does indeed have the observed $b(N)$ cut-off.

\section{RESULTS}
\label{sec:results}

\begin{figure*}
\resizebox{\textwidth}{!}{\includegraphics{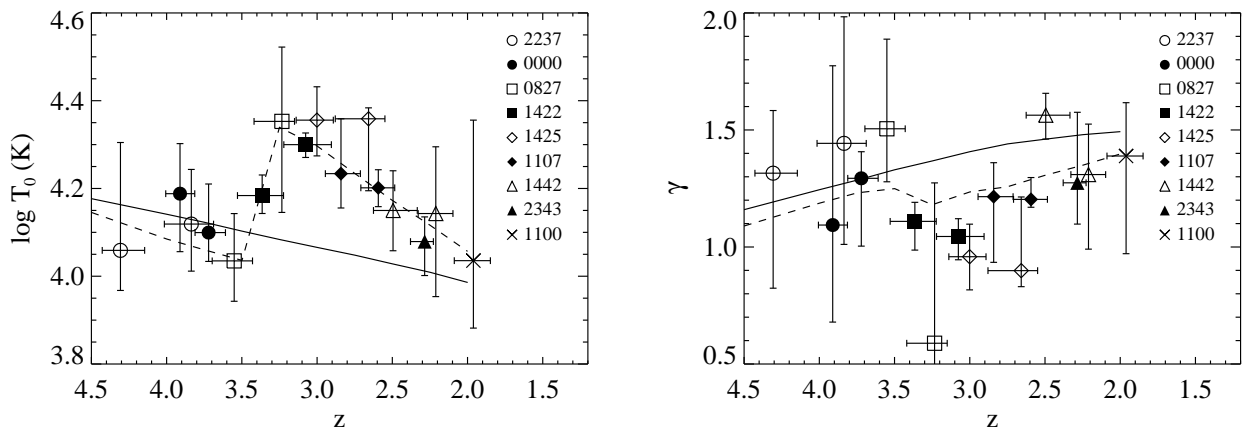}}
\caption{The temperature at the mean density (a) and the the slope of
the effective equation of state (b) as a function of
redshift. Horizontal error bars indicate the redshift interval spanned by
the absorption lines, vertical error bars are 1$\,\sigma$ errors. The
solid lines are for our reference model, L1, which uses the HM
ionizing background. The dashed lines are for a simulation that was
designed to fit the data. In this model, which has a much smaller
contribution from quasars at high redshift, \HeII\ reionizes at $z\sim
3.2$. Radiative transfer effects on the temperature of the IGM were
modeled schematically by increasing the photoheating rates for an
optically thin gas during reionization by a factor 4 for \HI\ and
\HeII\ and a factor 2 for \HeI. Although it is clear that the
temperature peaks at $z\sim 3$ and that the gas becomes close to
isothermal ($\gamma \sim 1$), the present constraints are not
sufficient to distinguish between a sharp rise (as indicated by the
dashed line) and a more gradual increase. } 
\label{fig:results}
\end{figure*}

The measured evolution of the temperature at the mean density and the
slope of the effective equation of state are plotted in
Fig.~\ref{fig:results}. From $z\sim 4$ to $z\sim 3$, $T_0$
increases and the gas becomes close to isothermal ($\gamma \sim
1.0$). This behavior differs drastically from that predicted by models
in which helium is fully reionized at higher redshift. For example,
the solid curves correspond to our reference model, L1, which uses a
uniform metagalactic UV-background from quasars as computed by HM and
which assumes the gas to be optically thin. In this simulation, both
hydrogen and helium are fully reionized by $z\sim 4.5$ and the
temperature of the IGM declines slowly as the universe expands. Such a
model can clearly not account for the peak in the temperature at
$z\sim 3$ (reduced $\chi^2$ for the solid curves are 6.8 for $T_0$ and
3.6 for $\gamma$). Instead, we associate the peak in $T_0$ and the low
value of $\gamma$ with reheating due to the second reionization of
helium (\HeII\ $\rightarrow$ \HeIII).

If reionization of \HeII\ happens
locally on a timescale that is short compared to the recombination
timescale, which for \HeIII\ is of the order of the age of the
universe at $z\sim 3$, then the energy density injected by
photoionization will be proportional to the gas density. Consequently,
the temperature increase will be independent of the density and the
equation of state of the IGM will become more isothermal. The change
in the slope of the equation of state at $z\sim 3$ is thus physically
consistent with our interpretation of the peak in the temperature at
the same redshift.

The dashed lines in Fig.~\ref{fig:results} are for a model that was
constructed to fit 
the data (reduced $\chi^2$ is 0.22 for $T_0$ and 1.38 for
$\gamma$). This model, for which stellar sources ionize \HI\ and
\HeI\ by $z\sim 5$ and quasars ionize \HeII\ at $z\sim 3.2$, has a
much softer spectrum at high redshift. Before
reionization, when the gas is optically thick to ionizing photons, the
mean energy per photoionization is much higher than in the optically
thin limit \cite{abel99:rad_transfer_igm}. We have approximated this
effect in this simulation by enhancing the photoheating rates during
reionization, so raising the temperature of the IGM.

Since the simulation assumes a uniform ionizing background, the
temperature has to increase abruptly (i.e.\ much faster than the gas
can recombine) in order to make $\gamma$ as small as observed. In
reality, the low-density gas may be reionized by harder photons, which
will be the first ionizing photons to escape from the dense regions
surrounding the sources. This would lead to a larger temperature
increase in the more dilute, cooler regions, resulting in a decrease
of $\gamma$ even for a more gradual reionization. Furthermore,
although reionization may proceed fast locally (as in our small
simulation box), it may be patchy and take some time to
complete. Hence the steep temperature jump indicated by the dashed
line, although compatible with the data, should be regarded as
illustrative only. The globally averaged $T_0$ could well increase
more gradually which would also be consistent with the data. 

Note that if the reionization is patchy, it would
give rise to large spatial fluctuations in the temperature. 
Because we measure the temperature-density relation from 
the lower cut-off of the $b(N)$-distribution, our results should be
regarded as lower limits to the average temperature. Absorption lines
arising in a local, hot ionization bubble would not necessarily raise
the observed cut-off in the $b(N)$-distribution.

The errors in Fig.~\ref{fig:results} can be directly
traced back to the corresponding $b(N)$-distributions
(Fig.~\ref{fig:mosaic}). Take for example 0827a, which has an
extremely low value of $\gamma$, with a very large error. This is
clearly due to the gap in the $b$-distribution at $\log N\sim
12.5$--13.0. The lack of lines in that region could be a statistical
fluctuation, or it could be an indication of large variations in the
temperature of the IGM.

\section{SYSTEMATICS}
\label{sec:systematics}

In this section we will investigate whether there are any systematic
effects that could affect our results.

\subsection{Numerical resolution}

\begin{figure*}
\resizebox{0.7\textwidth}{!}{\includegraphics{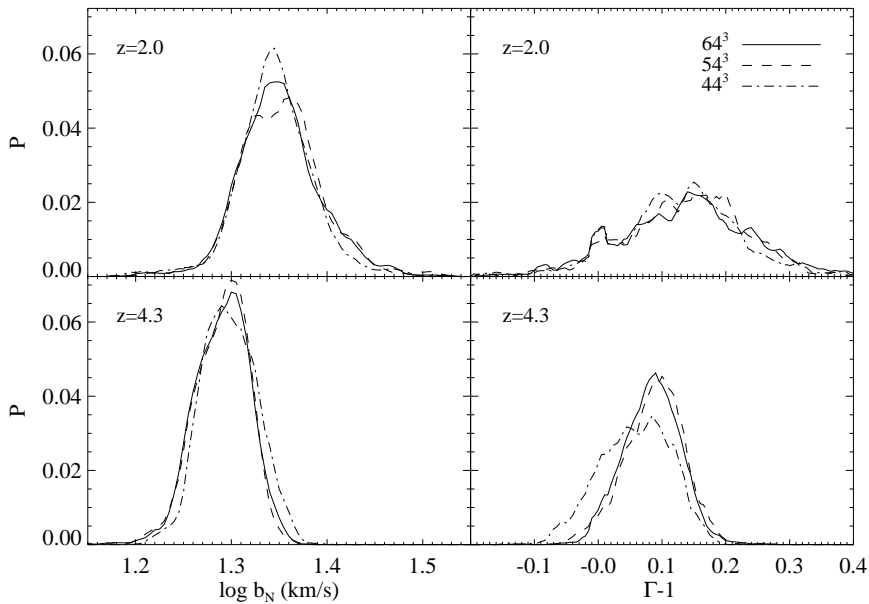}}
\caption{The effect of numerical resolution on the $b(N)$
cut-off. Shown are the probability distributions for the intercept
(left) and slope (right) of the cut-off in the $b(N)$-distributions of
models L1-Q1100 ($\bar{z}=1.97$) (top) and L1-Q2237b ($\bar{z}=4.27$)
(bottom), using $2\times 64^3$ (solid), $2\times 54^3$ (dashed) and
$2\times 44^3$ (dot-dashed) particles respectively. The small
difference between the intermediate and high resolution simulations
indicates that the latter has converged. The intercept was measured at
a column density of $10^{14.0}\,\cm^{-2}$ for L1-Q2237b and
$10^{13.4}\,\cm^{-2}$ for L1-Q1100.}
\label{fig:resolution}
\end{figure*}

The $b$-distribution has been shown to be very sensitive to numerical
resolution~\cite{theuns98:apmsph,bryan99:lya_numeffects}. It is
therefore important to check that the lower limit to the line widths
in our simulations is not set by the numerical resolution. We have
resimulated our reference model, L1, which is the second coldest
model, twice more at a lower resolution. The resolution was decreased
by decreasing the number of particles from $2\times 64^3$ to $2\times
54^3$ and $2\times 44^3$ respectively, while keeping the size of the
simulation box constant. The resulting probability distributions for
the intercept and the slope of the $b(N)$ cut-off are plotted in
Fig.~\ref{fig:resolution} for our lowest and highest redshift
samples. Only for the lowest resolution simulation do the differences
become noticeable. The intercept increases slightly and the slope
becomes slightly shallower, indicating that the lines at the low
column density end are not resolved. We conclude that the simulations
used for this work have sufficient resolution to provide an accurate
determination of the effective equation of state of the IGM.

\subsection{Simulation box size}

In order to investigate the effect of the simulation box size, we
resimulated model~L3 using a larger box, but with the same resolution
($5\,\hmpc$ and $2\times 128^3$ particles instead of $2.5\,\hmpc$ and
$2\times 64^3$ particles).  The effect of increasing the size of the
simulation box is more difficult to determine than the effect of
numerical resolution. When the box size is increased, the gas becomes
slightly hotter ($T_0$ increases by a few per cent), presumably
because shock heating is more effective due to the larger infall
velocities. This effect is small and since it does not affect the
relation between the $b(N)$ cut-off and the equation of state, it is
unimportant for this work. We find that using the larger simulation
box results in higher values of $T_0$. The effect is negligible at
$z\sim 4$ and increases to $\Delta \log T_0 \approx 0.05$ ($\Delta
T_0/T_0 \approx 0.12$) at $z\sim 2$. The change in the derived value
of $\gamma$ is very small ($\la 0.05$). Although $T_0$ and $\gamma$
may change a bit more if we increase the box size further, the small
difference between the two box sizes indicates that the effect of the
box size is insignificant compared to the statistical errors.

\subsection{Cosmology}

STLE showed that the relation between the cut-off in the $b(N)$
distribution and the equation of state is independent of the assumed
cosmology. However, the initial power spectra of the models
investigated by STLE were all normalized to match the observed
abundance of galaxy clusters at $z=0$. Bryan \&
Machacek~\shortcite{bryan00:cutoff} claimed that the $b$-distribution
depends strongly on the amplitude of the power spectrum, as predicted
by the model of Hui \& Rutledge~\shortcite{hui99:bdistr}. However,
Theuns et al.~\shortcite{theuns00:broadening} found the dependence to
be very weak, provided that the line fitting is done using an
algorithm, like \vpfit, that attempts to deblend absorption lines into
a set of thermally broadened components. Although the absorption
features do become broader for models with less small-scale power, the
curvature in the line centers does not change much. Consequently, the
total number of Voigt profile components used by \vpfit\ will
generally increase, but the fits to the line centers will change very
little.

To quantify the effect of decreasing the amount of small-scale power,
we resimulated our reference model~L1 twice using a lower
normalization of the initial power spectrum. We find that normalizing
to $\sigma_8=0.65$ instead of $\sigma_8=0.9$, changes the derived
values of $T_0$ by less than 3 per cent for all redshifts. The change
in $\gamma$ is never greater than 0.1. For the extreme case of
$\sigma_8=0.4$, the derived values of $T_0$ are about 15 per cent
lower and $\gamma$ differs by about 0.15. We conclude that the effect
of the uncertainty in the normalization of the primordial power
spectrum is small.

\subsection{Continuum fitting and the mean absorption}

Errors in the continuum fit of the observed spectra, will lead to
errors in the effective optical depth. Underestimating the observed
continuum will decrease the measured effective optical depth.
Decreasing the mean absorption in the simulations will increase the
density corresponding to a given column density. Hence the slope of
the cut-off will remain unchanged, but the intercept will increase,
although the effect is small (STLE). Increasing the intercepts of the
calibrating simulations will decrease the derived temperature
$T_\delta$. The higher derived density contrast will work in the same
direction (provided that $\gamma > 1$), resulting in a lower $T_0$.

The measured effective optical depth for sample 2343 seems to be
relatively high compared to the other samples
(Fig.~\ref{fig:tau_eff}). We tried scaling the synthetic spectra to
the effective optical depth corresponding to the dashed line in
Fig.~\ref{fig:tau_eff} at the redshift of 2343, which is about 30 per
cent lower than the measured value. This resulted in an increase of
$\log T_0$ by 0.04 ($\Delta T_0/T_0 \approx 0.09$), while leaving
$\gamma$ unchanged.

In addition to errors in the continuum fit of the observed spectra,
the continuum fitting of the synthetic spectra could also lead to
systematic errors. We checked this by repeating the analysis of the
simulations corresponding to our highest redshift sample, 2237b, but
this time without continuum fitting the synthetic spectra. In this
case the derived value of $T_0$ would be 9 per cent lower, while the
value of $\gamma$ would be higher by 0.07. Hence systematic effects in
the continuum fitting of the observed and the synthetic spectra are
unlikely to be important.

\section{SUMMARY AND DISCUSSION}
\label{sec:discussion}
We have measured the cut-off in the distribution of line widths ($b$)
as a function of column density ($N$) in a set of nine high-quality
\lya\ forest spectra, spanning the redshift range 2.0--4.5. We
emphasized that the evolution of the temperature of the intergalactic
medium (IGM) cannot be derived directly from the evolution of the
$b(N)$ distribution, the decrease of the overdensity corresponding to
a fixed column density with redshift has to be taken into account. We
therefore used hydrodynamic simulations to calibrate the relations
between the $b(N)$ cut-off and the temperature-density relation of the
low density gas. The calibration was done separately for each observed
spectrum, using synthetic spectra that were processed to give them
identical characteristics as the observed spectrum. Crucially, Voigt
profiles were fitted to the real and simulated spectra using the same
automated fitting package (a modified version of \vpfit). We have
checked possible systematic errors arising from the finite numerical
resolution, the finite size of the simulation box, the amplitude of
the initial power spectrum and continuum fitting errors in both
synthetic and observed spectra. In all cases, the effects were small
compared to the statistical errors.

The measured thermal evolution differs drastically from the scenario
predicted by current models of the ionizing background from quasars,
in which helium is fully reionized by $z\sim 4.5$. The temperature at
the mean density, $T_0$, increases from $z\sim 4$ to $z\sim 3$, after
which it decreases again (Fig.~\ref{fig:results}). The slope of the
equation of state reaches a minimum at $z\sim 3$, where it becomes
close to isothermal. More data at $z \ga 3$ is needed to determine
whether the rise in $T_0$ is sharp or gradual. These results suggest
that the low density IGM was reheated from $z\sim 4$--3, which we
interpret as reheating associated with the reionization of \HeII.

These results are in qualitative agreement with those reported
recently by Ricotti et al.~\shortcite{ricotti00:eos}. They used a
method which relies on the assumption that only thermal broadening
contributes to the line widths of the absorption lines at the peak of
the $b$-distribution and used approximate simulation techniques to
determine the density - column density relation. By applying their
method to published lists of Voigt profile fits they found that at
$z\sim 3$, $\gamma$ is smaller than would be expected if the
reionization of helium had been completed at high redshift. Ricotti et
al.\ have only three data points in the range $z=2$--4 which all
overlap at the 0.5 sigma level, so we cannot compare the shape of the
temperature evolution. However, it is interesting that they measure a
temperature at $z=2$ and $z=4$ that is about seventy per cent higher
than is reported here, although their error bars are sufficiently
large to agree with our results at the 1$\sigma$ level. Since Jeans
smoothing contributes to the line widths, as
Fig.~\ref{fig:intercept-logt} indicates (see also Theuns et al.~2000),
their method should lead to an overestimate of the temperature (by
about sixty per cent for the example in
Fig.~\ref{fig:intercept-logt}). The contribution of Jeans smoothing to
the line width may be larger for the lines near the peak of the
$b$-distribution than indicated in Fig.~\ref{fig:intercept-logt},
which is for the narrower lines near the cut-off.
 
Although the reionization of helium may not be the only process which can
explain the peak in $T_0$ at $z\sim 3$, it appears to be the only
process that can simultaneously account for the observed decrease in
$\gamma$. Galactic winds for example, would be less important
in the low density regions, and would therefore result in an increase
in the slope of the equation of state. Another possible explanation is
a hardening of the ionizing background at $z\sim 3$, as might be
expected because of the increase in the number of quasars. If helium
had already been ionized, this would still raise the temperature
somewhat, but the effect would be stronger in the high density regions
where the gas recombines faster. Hence this would also lead to an
increase in $\gamma$, contrary to what is observed.

Recently, it was shown~\cite{theuns98:apmsph,bryan99:lya_numeffects}
that high resolution simulations of the standard cold dark model,
using the ionizing background computed by HM produce a larger fraction
of narrow lines than observed at $z\sim 3$.  Different authors have
proposed different solutions to this problem. Theuns et
al.~\shortcite{theuns98:apmsph} suggested that the gas temperature in
the simulations was too low, while Bryan et
al.~\shortcite{bryan99:lya_numeffects} argued that the amplitude of
primordial fluctuations was too high.  For a given reionization
history, the temperature in the simulations could for example be
increased by increasing the baryon density and the age of the universe
\cite{theuns99:cosmology}, including Compton heating by the hard X-ray
background \cite{madau99:xray} and possibly photo-electric heating by
dust grains \cite{nath99:dust_heating}. A comparison of our reference
model L1, which uses the HM ionizing background (solid lines in
Fig.~\ref{fig:results}), with the data clearly shows that the
model underestimates the temperature at $z\sim
3$. Since the above mentioned mechanisms for increasing the
temperature do not change the overall shape of the thermal evolution,
a change in the reionization history is required to bring the data and
simulations into agreement. Furthermore, the photoheating rates around
reionization need to be enhanced to account for the fact that heating
by photons with energies significantly above the ionization potential
is important when the gas is not optically thin
\cite{abel99:rad_transfer_igm}, as is generally assumed in
cosmological simulations.

There are two other lines of evidence for late reionization of
\HeII. The first are direct measurements of the optical depth from
\HeII\ \lya\ absorption, which have so far been obtained for four
quasars
\cite{jakobsen94:he,davidsen96:he,reimers97:he,anderson99:he,heap00:he}.
These observations already provide strong evidence for a drop in the
mean absorption from $z\sim 3.0$ to 2.5. The second piece of evidence
concerns a change in the spectral shape of the ionizing background. As
\HeII\ is ionized, the mean free path of hard UV photons will increase
and the spectrum of the UV background will become harder. Songaila \&
Cowie \shortcite{songaila96:metals} and Songaila
\shortcite{songaila98:ionizing_background} have reported a rapid
increase with decreasing redshift of the \SiIV/\CIV\ ratio at $z\sim
3$, which they interpreted as evidence for a sudden reionization of
\HeII. However, Boksenberg et
al.~\shortcite{boksenberg98:ionizing_background} found only a gradual
change with redshift. The interpretation of this metal line ratio is
complicated because local stellar radiation is likely to be important
\cite{giroux97:siiv/civ}.

It should be kept in mind that these three different types of
observations probe different physical structures. Our results apply to
density fluctuations around the cosmic mean, the effective optical
depth depends mostly on the neutral fraction in the voids and the
metal line ratios probe the high density peaks. These structures will
probably not be ionized simultaneously. After the ionization front
breaks through the haloes surrounding the source of \HeII\ ionizing
photons, e.g.\ a quasar, it will propagate quickly into the voids. The
filaments, where the recombination rate is much higher, will get
ionized more slowly, starting from the outside
\cite{miralda-escude00:reionization,gnedin00:reionization}.  The IGM
will still be thick to \lya\ photons when only a small neutral
fraction remains in the voids. Hence a drop in the \HeII\ \lya\
optical depth at $z\sim 3$ would suggest that the reionization of the
voids, which cover most of the volume, is complete
\cite{miralda-escude98:reionization}. It is important to note that the
peak in $T_0$ at $z\sim 3$ does not imply that the temperature of the
general IGM reaches a maximum. In fact, our results imply that the
temperature of slightly overdense gas ($\delta \ga 2$) is almost
constant because the slope of the equation of state, $\gamma$, is
minimum when $T_0$ is maximum.

Detailed modeling, probably in the form of large hydrodynamical
simulations, which include radiative transfer, is required to see
whether the various observational constraints can be fit into a
consistent picture. However, the ingredients necessary to explain our
discovery of a peak in the temperature of the low density IGM at
$z\sim 3$ are clear even from our crude model: a softer background at
high redshift to delay helium reionization and enhanced heating rates
compared to the optically thin limit. Once the evolution of helium
heating is understood, the measurements of the temperature at higher
redshifts can be used to constrain the epoch of hydrogen reionization.

Finally, we would like to note that because of their hard spectrum,
quasars tend to ionize helium shortly after hydrogen, although the
delay depends on the clumpiness of the IGM
\cite{madau98:reionization}. It may therefore be difficult to postpone
the reionization of helium until $z\sim 3$, if quasars were
responsible for reionizing hydrogen at $z> 5$.  Hence the mounting
evidence for helium reionization at $z\sim 3$ suggests that hydrogen
was reionized by stars.

\section*{ACKNOWLEDGMENTS}
We would like to thank Bob Carswell for letting us use the spectrum of
Q1100$-$264 and for helping us with \vpfit. We are also grateful to
Sara Ellison for letting us use the HIRES spectra of
APM~08279+5255. We thank Martin Haehnelt, Lam Hui, Jordi
Miralda-Escud\'e and Martin Rees for stimulating discussions. JS
thanks the Isaac Newton Trust, St.~John's College and PPARC for
support, WLWS acknowledges support from NSF under grant AST-9900733
and GE thanks PPARC for the award of a senior fellowship. Research was
conducted in cooperation with Silicon Graphics/Cray Research utilising
the Origin 2000 supercomputer COSMOS, which is a UK-CCC facility
supported by HEFCE and PPARC. This work has been supported by the TMR
network on `The Formation and Evolution of Galaxies', funded by the
European Commission.

{}

\end{document}